\documentclass[10pt,a4paper]{article}

\usepackage[latin1]{inputenc} 
\usepackage[dvips]{graphicx} 
\usepackage{a4wide}
\usepackage{latexsym}
\usepackage{amsmath}
\usepackage{amssymb}
\usepackage{dsfont}
\usepackage{color}

\title{Electromagnetic Properties for Arbitrary Spin Particles:\\
Part 1 $-$ Electromagnetic Current and Multipole Decomposition}

\author{Cédric Lorcé\\ \small{\emph{Institut für Kernphysik, Johannes Gutenberg-Universität, D-55099 Mainz, Germany}}\\
\small{\emph{E-mail: lorce@kph.uni-mainz.de}}}
\date{}

\newcommand{\ud}{\mathrm{d}}

\newcommand{\pslash}{p\!\!\!/}

\begin{document}

\maketitle

\begin{center}
\begin{minipage}[t]{15cm}
\small{In a set of two papers, we propose to study an old-standing problem, namely the electromagnetic interaction for particles of arbitrary spin. Based on the assumption that light-cone helicity at tree level and $Q^2=0$ should be conserved non-trivially by the electromagnetic interaction, we are able to derive \emph{all} the natural electromagnetic moments for a pointlike particle of \emph{any} spin. In this first paper, we propose a transparent decomposition of the electromagnetic current in terms of covariant vertex functions. We also define in a general way the electromagnetic multipole form factors, and show their relation with the electromagnetic moments. Finally, by considering the Breit frame, we relate the covariant vertex functions to multipole form factors.}
\end{minipage}
\end{center}

\section{Introduction}

High-energy physics involves high-spin particles for many reasons. Among those, let us mention that:
\begin{itemize}
\item experimentally, more than fifty baryons with spins ranging from $3/2$ to $15/2$ have been observed \cite{PDG};
\item these resonances enter as intermediate states in the description of the photo- and electro-pion production off protons, which are a main focus of experiments at electron facilities such as Jefferson Lab, ELSA, MAMI \cite{MAMI};
\item in proposals for physics beyond the Standard Model based on supersymmetry, which will be explored soon by the LHC, elementary particles with high spin (\emph{e.g.} the gravitino) are required. 
\end{itemize}
For further motivations to study high-spin particles, see \emph{e.g.} \cite{ref} and references therein.

Formalisms to study high-spin particles have been proposed a long time ago. Dirac, Fierz and Pauli (DFP) were the first to develop a general theory of particles with arbitrary spin \cite{DFP}, but this formalism is quite complicated. Later, Rarita and Schwinger (RS) proposed a more convenient formalism for arbitrary half-integer spin particles \cite{RS}, which is still the most often used to date in the literature. In the case of integer spin particles, the Klein-Gordon (KG) equation with subsidiary conditions is more suited. Actually, it has been shown by Moldauer and Case that both RS and KG approaches can be derived from the DFP formalism \cite{gyro}. There exists another formalism due to Bargmann and Wigner \cite{BW} from which one can also derive the RS and KG equations together with their subsidiary conditions \cite{eq}.

Theories with spin $>1$ are however plagued by arbitrary parameters in both their Lagrangian and propagator \cite{gyro,disease}. This is due to the fact that the corresponding fields contain extra degrees of freedom related to lower spins. To eliminate these extra degrees of freedom, one imposes constraints (the subsidiary conditions in RS and KG formalisms). Unfortunately, interactions usually break these constraints, generating many problems. Among the various difficulties \cite{difficulties}, let us mention an important result of Velo and Zwanziger \cite{VZ} that showed that a massive $c$-number spin-$3/2$ field minimally coupled to an external electromagnetic field will lead to the existence of tachyons, and therefore to non-causality. The causality issue has been shown to be related to the value of the gyromagnetic factor at tree level in the case of elementary spin-$3/2$ particles \cite{ref,Ferrara}.

The gyromagnetic factor of elementary particles has also a long history. Based on the minimal substitution $p^\mu\to\pi^\mu\equiv p^\mu-e\,A^\mu$, it was argued that the gyromagnetic factor at tree level depends on the spin $j$ of the particle \cite{gyro,Belifante} as
\begin{equation}\label{belifante}
g_j=\frac{1}{j},
\end{equation}
\emph{i.e.} the magnetic moment depends only on the mass and electric charge of the particle. This result agrees with the case $j=1/2$ but seems to disagree with higher-spin cases. Many reasons suggest that the gyromagnetic factor at tree level is in fact \emph{independent} of the spin \cite{indep}
\begin{equation}
g_j=2,
\end{equation}
\emph{i.e.} the magnetic moment is directly proportional to the spin. Among the various reasons, let us mention that
\begin{itemize}
\item the only higher-spin and charged elementary particle observed is the $W$ boson. At tree level, it has $g_W=2$ \cite{ExpSM} and not $g_W=1$ as suggested by \eqref{belifante} for spin-1 particles;
\item the relativistic equation of motion of the polarization four-vector $S_\mu$ in a homogeneous external electromagnetic field is simplified for $g_j=2$ \cite{poleq};
\item in the supersymmetric sum rules framework, when all magnetic-moment matrix elements are diagonal, the gyromagnetic factor of arbitrary-spin supersymmetric
particles must be equal to 2 \cite{SUSY};
\item string theory also suggests that $g_j=2$ \cite{string};
\item by requiring that forward Compton scattering amplitudes of physical theories possess a good high-energy behavior, Weinberg showed \cite{unitarity} that this implies $g_j\approx 2$ for any non-strongly interacting spin-$j$ particle.
\end{itemize}
The way to reconcile Lagrangian theories with $g_j=2$ is to allow non-minimal coupling \cite{NMC}. Indeed, using the minimal substitution is ambiguous for spins $j>1/2$ since one has $\left[\pi^\mu,\pi^\nu\right]=-ie\,F^{\mu\nu}$ while $\left[p^\mu,p^\nu\right]=0$. Note that a non-minimal coupling $F^{\mu\nu}W_\mu^+W_\nu^-$ is already present in the Standard Model \cite{SM}.

In order to characterize completely the electromagnetic interaction of an elementary particle with spin $j>1/2$, the electric charge and the magnetic dipole moment are not sufficient. In general, a spin-$j$ particle has $2j+1$ electromagnetic multipoles if Lorentz, parity and time-reversal symmetries are respected. The knowledge of these ``natural'' electromagnetic moments is obviously very important since they
\begin{itemize}
\item constrain the construction of a consistent higher-spin electromagnetic interaction theory;
\item shed light on the relation between causality and unitarity in the ultrarelativistic limit, as suggested already by the gyromagnetic factor;
\item allow a better understanding of internal structure of composite particles with spin $j>1/2$;
\item \ldots
\end{itemize}

In a set of two papers, we present our results concerning the electromagnetic interaction for particles with arbitrary spin. Based on the simple assumption that QED conserves non-trivially the light-cone helicity of \emph{any} elementary particle at tree level and real photon point $Q^2=0$, we were able to derive \emph{all} natural electromagnetic moments for particles of \emph{any} spin. 

This paper contains the first part of our investigations. In Section \ref{current}, we give the general form of the electromagnetic current in terms of covariant vertex functions. An explicit multipole decomposition is then performed in Section \ref{decomposition} together with a transparent relation to electromagnetic moments. In Section \ref{BF}, we use the Breit frame to obtain a general relation between multipole form factors and covariant vertex functions.

\section{Electromagnetic Current for Arbitrary Spin}\label{current}

The interaction of a particle with the electromagnetic field can be described in terms of matrix elements, also known as on-shell vertex functions, of the following form
\begin{equation}\label{Qcurrent}
J^\mu\equiv\langle p',\lambda'|J_\textrm{EM}^\mu(0)|p,\lambda\rangle,
\end{equation}
where $p$ (resp. $p'$) is the initial (resp. final) four-momentum of the particle, and $|p,\lambda\rangle$ is the spin-$j$ single-particle state with four-momentum $p$ and polarization $\lambda$. For further convenience, we define the scalar quantity $Q^2\equiv-q^2$ as \emph{minus} the square of the four-momentum transfer $q=p'-p$ due to the photon. 

If the particle respects parity and time-reversal symmetries, its matrix elements can in principle be conveniently written in terms of $2j+1$ independent covariant vertex functions $F_k(Q^2)$ \cite{ind}, also often called form factors. Note however that the decomposition is not unique. In this sense, the matrix elements are more fundamental than the covariant vertex functions.

Since any spin representation can be constructed from spin-$1/2$ and spin-$1$ representations, it is rather straightforward to obtain the general structure of the matrix elements satisfying Lorentz covariance, gauge invariance, together with parity and time-reversal symmetries. We will use the convenient RS and KG formalisms for the description of the particle polarizations.

\subsection{Arbitrary Integer Spin Current}

A particle with integer spin $j$ and mass $M$ can be described in terms of a polarization
tensor $\varepsilon_{\alpha_1\cdots\alpha_j}(p,\lambda)$. This tensor is
completely symmetric and satisfies subsidiary conditions, namely transversity and tracelessness
\begin{equation}\label{subs}
\begin{split}
p^\mu \varepsilon_{\mu\alpha_2\cdots\alpha_j}(p,\lambda)=&\,0,\\
\varepsilon^\mu_{\mbox{ }\mu\alpha_3\cdots\alpha_j}(p,\lambda)=&\,0,
\end{split}
\end{equation} 
in order to insure the correct number of degrees of freedom $2j+1$.

The electromagnetic interaction current for
integer spin particles has the following general simple form
\begin{equation}\label{intcurr}
J^\mu=(-1)^j\,\varepsilon^*_{\alpha'_1\cdots\alpha'_j}(p',\lambda')\left[P^\mu\sum_{(k,j)}F_{2k+1}(Q^2)+\left(g^{\mu\alpha_j}q^{\alpha'_j}-g^{\alpha'_j\mu}q^{\alpha_j}\right)\sum_{(k,j-1)}F_{2k+2}(Q^2)\right]\varepsilon_{\alpha_1\cdots\alpha_j}(p,\lambda),
\end{equation}
where
\begin{equation}
\sum_{(k,j)}\equiv\sum_{k=0}^j\left[\prod_{i=1}^k\left(-\frac{q^{\alpha'_i}q^{\alpha_i}}{2M^2}\right)\prod_{i=k+1}^j g^{\alpha'_i\alpha_i}\right],
\end{equation}
and $P=p'+p$ is twice the averaged four-momentum of the particle.

To illustrate this, let us consider for example the case $j=1$. The explicit expression of \eqref{intcurr} becomes
\begin{displaymath}
J^\mu_{(1)}=-\varepsilon^*_{\alpha'}(p',\lambda')\left[P^\mu\,g^{\alpha'\alpha}\,F_1(Q^2)+\left(g^{\mu\alpha}q^{\alpha'}-g^{\alpha'\mu}q^\alpha\right)F_2(Q^2)-P^\mu\,\frac{q^{\alpha'}q^\alpha}{2M^2}\,F_3(Q^2)\right]\,\varepsilon_\alpha(p,\lambda),
\end{displaymath}
which coincides with the well-known expression for the electromagnetic interaction of vector particles \cite{vect} provided that
\begin{displaymath}
F_k(Q^2)=G_k(q^2),\qquad k=1,2,3.
\end{displaymath}

\subsection{Arbitrary Half-Integer Spin Current}

A particle with half-integer spin $j=n+\frac{1}{2}$ and mass $M$ can be described in terms of a polarization
spin-tensor $u_{\alpha_1\cdots\alpha_n}(p,\lambda)$.
This spin-tensor, besides being completely symmetric in its Lorentz indices, has to fulfill a set of equations\footnote{Note that the usual subsidiary conditions of integer spin \eqref{subs} applied to the spin-tensor are just consequences of \eqref{RScond}. Indeed, they can be obtained by multiplying the two equations in \eqref{RScond} on the left by $\gamma^{\alpha_1}$ and $\gamma^{\alpha_2}$, respectively.} \cite{RS}
\begin{equation}\label{RScond}
\begin{split}
\left(\pslash-M\right)u_{\alpha_1\cdots\alpha_n}(p,\lambda)=&\,0,\\
\gamma^\mu u_{\mu\alpha_2\cdots\alpha_n}(p,\lambda)=&\,0,
\end{split}
\end{equation}
in order to insure the correct number of degrees of freedom $2j+1$.

The electromagnetic interaction current for
half-integer spin particles has the following general simple form
\begin{equation}\label{halfintcurr}
J^\mu=(-1)^n\,\bar u_{\alpha'_1\cdots\alpha'_n}(p',\lambda')\sum_{(k,n)}\left[\gamma^\mu\,F_{2k+1}(Q^2)+\frac{i\sigma^{\mu\nu}q_\nu}{2M}\,F_{2k+2}(Q^2)\right]u_{\alpha_1\cdots\alpha_n}(p,\lambda).
\end{equation}

To illustrate this, we will consider two examples. Let us consider first the case $j=1/2$. The explicit expression of \eqref{halfintcurr} becomes
\begin{displaymath}
J^\mu_{(1/2)}=\bar
u(p',\lambda')\left[\gamma^\mu\,F_1(Q^2)+\frac{i\sigma^{\mu\nu}q_\nu}{2M}\,F_2(Q^2)\right]u(p,\lambda),
\end{displaymath}
which coincides with the well-known expression for the electromagnetic interaction of spin-$1/2$ particles. The covariant vertex functions $F_1(Q^2)$ and $F_2(Q^2)$ are known in the literature as the Dirac and Pauli form factors, respectively. Let us consider now the case $j=3/2$. The explicit expression of \eqref{halfintcurr} becomes
\begin{displaymath}
J^\mu_{(3/2)}=-\bar u_{\alpha'}(p',\lambda')\left\{g^{\alpha'\alpha}\left[\gamma^\mu\,F_1(Q^2)+\frac{i\sigma^{\mu\nu}q_\nu}{2M}\,F_2(Q^2)\right]-\frac{q^{\alpha'}q^\alpha}{2M^2}\left[\gamma^\mu\,F_3(Q^2)+\frac{i\sigma^{\mu\nu}q_\nu}{2M}\,F_4(Q^2)\right]\right\}\,u_\alpha(p,\lambda),
\end{displaymath}
which coincides with the well-known expression for the electromagnetic interaction of spin-$3/2$ particles \cite{vdh} provided that
\begin{displaymath}
\begin{split}
F_1(Q^2)&=\,F_1^*(Q^2)=a_1(q^2)+a_2(q^2),\\
F_2(Q^2)&=\,F_2^*(Q^2)=-a_2(q^2),\\
F_3(Q^2)&=\,-\frac{1}{2}\,F_3^*(Q^2)=-\frac{1}{2}\left[c_1(q^2)+c_2(q^2)\right],\\
F_4(Q^2)&=\,-\frac{1}{2}\,F_4^*(Q^2)=\frac{1}{2}\,c_2(q^2).
\end{split}
\end{displaymath}
As it will become obvious in the following, the decompositions \eqref{intcurr} and \eqref{halfintcurr} we propose are such that spurious $(-\frac{1}{2})$ factors are avoided. Note also that the electromagnetic gauge invariance is explicit, \emph{i.e.} one can easily see that $q_\mu J^\mu=0$.

\subsection{Polarization Tensor and Spin-Tensor}

In order to work out the vertex function in specific cases, we need an explicit form for the polarization \mbox{(spin-)tensor}. The easiest way has been proposed a long time ago by Auvil and Brehm \cite{eq,pol}. By always coupling the maximum possible spin of a lower-order polarization tensor and a polarization four-vector, the product will satisfy KG or RS equations together with the subsidiary conditions. One therefore obtains the following recursion formula 
\begin{equation}\label{tensorprod}
T_{\alpha_1\cdots\alpha_n}(p,\lambda)=\sum_{m,m'}\langle
1m,(j-1)m'|j\lambda\rangle\,\varepsilon_{\alpha_1}(p,m)\,T_{\alpha_2\cdots\alpha_n}(p,m'),
\end{equation}
where $\langle
j_1m_1,j_2m_2|jm\rangle$ represents the Clebsch-Gordan coefficient in the Condon-Shortley phase convention, and $T_{\alpha_1\cdots\alpha_n}(p,\lambda)$ stands for the polarization spin-tensor $u_{\alpha_1\cdots\alpha_n}(p,\lambda)$ with $j=n+\frac{1}{2}$  (half-integer spin case)
or the polarization tensor $\varepsilon_{\alpha_1\cdots\alpha_n}(p,\lambda)$ with $j=n$ (integer spin case). For the spin-tensor of half-integer spin, it is however more convenient to consider
\begin{displaymath}
u_{\alpha_1\cdots\alpha_n}(p,\lambda)=\sum_{m,m'}\langle
\frac{1}{2}\frac{m}{2},nm'|j\lambda\rangle\,u(p,m)\,\varepsilon_{\alpha_1\cdots\alpha_n}(p,m'),
\end{displaymath}
which is naturally equivalent to \eqref{tensorprod}. More explicitly,
\begin{equation}\label{spintensor}
u_{\alpha_1\cdots\alpha_n}(p,\lambda)=\sqrt{\frac{j+\lambda}{2j}}\,u(p,+)\,\varepsilon_{\alpha_1\cdots\alpha_n}(p,\lambda-\frac{1}{2})+\sqrt{\frac{j-\lambda}{2j}}\,u(p,-)\,\varepsilon_{\alpha_1\cdots\alpha_n}(p,\lambda+\frac{1}{2}).
\end{equation}

We can therefore concentrate on the polarization tensor of integer spin. By interating \eqref{tensorprod}, one obtains
\begin{displaymath}
\varepsilon_{\alpha_1\cdots\alpha_j}(p,\lambda)=\sum_{m_l=0,\pm1}\left[\prod_{l=1}^{j-1}\langle1m_llm'_l|(l+1)m'_{l+1}\rangle\,\varepsilon_{\alpha_l}(p,m_l)\right]\varepsilon_{\alpha_j}(p,m_j),
\end{displaymath} 
where the sum is implicitly restricted to configurations such that $\sum_{l=1}^jm_l=\lambda$, and where
\begin{displaymath}
m'_l=m_j+\sum_{k=1}^{l-1} m_k.
\end{displaymath}
The Clebsch-Gordan coefficients can be written as \cite{angular}
\begin{displaymath}
\langle
1m_llm'_l|(l+1)(m_l+m'_l)\rangle=\sqrt{\frac{C_2^{1+m_l}C_{2l}^{l+m'_l}}{C_{2l+2}^{l+m_l+m'_l+1}}},
\end{displaymath}
where $C_n^k$ is the binomial function
\begin{equation}
C_n^k=\left(\begin{array}{c}n\\ k\end{array}\right)\equiv\left\{\begin{array}{cl} \frac{n!}{k!\,(n-k)!},&n\geq k\geq 0\\ 0,& \textrm{otherwise}\end{array}\right..
\end{equation}
One easily obtains
\begin{displaymath}
\varepsilon_{\alpha_1\cdots\alpha_j}(p,\lambda)=\sum_{m_l=0,\pm1}\frac{\prod_{l=1}^j\sqrt{C_2^{1+m_l}}\,\varepsilon_{\alpha_l}(p,m_l)}{\sqrt{C_{2j}^{j+\lambda}}},
\end{displaymath}
which can be rewritten more conveniently as \cite{Chung}
\begin{equation}\label{poltensor}
\varepsilon_{\alpha_1\cdots\alpha_j}(p,j-m)=\quad\sum_{k=0}^{m/2}\frac{\sum_\mathcal{P}\left[\prod_{l=1}^k\varepsilon_{\alpha_{\mathcal{P}(l)}}(p,-)\right]\left[\prod_{l=k+1}^{m-k}\varepsilon_{\alpha_{\mathcal{P}(l)}}(p,0)\right]\left[\prod_{l=m-k+1}^j\varepsilon_{\alpha_{\mathcal{P}(l)}}(p,+)\right]}{2^{k-m/2}\,k!\,(m-2k)!\,(j-m+k)!\,\sqrt{C_{2j}^m}},
\end{equation}
where $\mathcal{P}$ stands for a permutation of $\{1,\cdots,j\}$. The presence of factorials in the denominator is due to the fact that permuting the indices of two polarization four-vectors with the same polarization does not give a new contribution.

\section{Electromagnetic Moments}\label{decomposition}

Our final aim is to obtain the natural values for the electromagnetic moments of a particle of arbitrary spin. These moments are related to the so-called multipole form factors $G_{El}(Q^2)$ and $G_{Ml}(Q^2)$ at real photon point $Q^2=0$. Thanks to an explicit multipole decomposition of a four-current $j^\mu$, we define all the magnetic and electric multipole form factors and present the explicit relation with the (cartesian) electromagnetic moments. 

For further convenience, we denote by $-e$ the electric charge of an electron, and we introduce the notation
\begin{equation}
\tau\equiv\frac{Q^2}{4M^2}.
\end{equation}

\subsection{Multipole Decomposition}

The zero component $\mu=0$ of a four-current $j^\mu$ corresponds to the charge density while the spatial components $\mu=i$ are related to some kind of ``magnetic density'' \cite{moments}. Let us refer, for the moment, to these densities by means of the generic density
\begin{equation}
\rho(\vec{q})=\int\ud^3r\,e^{i\vec{q}\cdot\vec{r}}\,\rho(\vec{r}).
\end{equation}
The multipole decomposition of this density can be written in the following form ($Q\equiv |\vec{q}|$)
\begin{equation}\label{multdec}
\rho(\vec{q})=\sum_{l=0}^{+\infty}\sum_{m=-l}^lQ^l\,F_{lm}(Q^2)\,Y_{lm}(\Omega_q),
\end{equation}
where $Y_{lm}(\Omega)$ are the usual spherical harmonics, $\Omega$ stands for the couple $(\theta,\phi)$. Thanks to the identity \cite{angular}
\begin{displaymath}
e^{i\vec{q}\cdot\vec{r}}=4\pi\sum_{l=0}^{+\infty}\sum_{m=-l}^li^l\,j_l(Qr)\,Y_{lm}(\Omega_q)\,Y^*_{lm}(\Omega_r),
\end{displaymath}
where $j_l(x)$ are the spherical Bessel functions, and using orthonormal relations among the spherical harmonics
\begin{displaymath}
\int\ud\Omega\,Y^*_{l'm'}(\Omega)\,Y_{lm}(\Omega)=\delta_{ll'}\,\delta_{mm'},
\end{displaymath}
we find
\begin{displaymath}
F_{lm}(Q^2)=\frac{4\pi\,i^l}{Q^l}\int\ud^3r\,j_l(Qr)\,Y_{lm}(\Omega_r)\,\rho(\vec{r}).
\end{displaymath}
Remembering that we are interested in the electromagnetic properties of particles, one can naturally consider that the electric and magnetic densities exhibit an azimuthal (or cylindrical) symmetry with respect to the quantization axis. Let us identify the spin quantization axis with the $z$-axis. Due to this symmetry, the multipole decomposition \eqref{multdec} reduces to $m=0$ components only
\begin{equation}\label{multdec2}
\rho(\vec{q})=\sum_{l=0}^{+\infty}Q^l\,F_{l0}(Q^2)\,Y_{l0}(\Omega_q).
\end{equation}

The spherical harmonics $Y_{l0}(\Omega_r)$ are related to cartesian multipole
structures $C_l(\vec{r})$ as follows
\begin{displaymath}
C_l(\vec{r})=l!\,r^l\,\sqrt{\frac{4\pi}{2l+1}}\,Y_{l0}(\Omega_r)=l!\,r^l\,P_{l0}(\cos\theta_r),
\end{displaymath}
\emph{i.e.} more explicitly
\begin{displaymath}
\begin{split}
C_0(\vec{r})&=\,1,\\
C_1(\vec{r})&=\,z,\\
C_2(\vec{r})&=\,3\,z^2-r^2,\\
C_3(\vec{r})&=\,15\,z^3-9\,r^2z,\\
&\,\,\:\vdots
\end{split}
\end{displaymath}
Moreover, to the
lowest order in $Q$, the spherical Bessel functions $j_l(Qr)$ can be
written as
\begin{displaymath}
j_l(Qr)=\frac{(Qr)^l}{(2l+1)!!}+\mathcal{O}(Q^{l+2}).
\end{displaymath}
From this, one can see that the electromagnetic moments are proportional to $F_{l0}(0)$
\begin{equation}
\int\ud^3r\,C_l(\vec{r})\,\rho(\vec{r})=(-i)^l\,l!\,(2l-1)!!\,\sqrt{\frac{2l+1}{4\pi}}\,F_{l0}(0).
\end{equation}

\subsection{Electric Multipoles}

Electric multipoles\footnote{If we follow more rigorously the literature terminology, they should be called Coulomb multipoles.} are obtained from the charge density $\rho(\vec{r})=j^0(\vec{r})$. We define the electric multipole form factors $G_{El}(Q^2)$ as follows
\begin{equation}\label{Edef}
G_{El}(Q^2)\equiv(-i)^l\,\frac{(2l-1)!!}{l!}\,\frac{(2M)^l}{e}\,\sqrt{\frac{2l+1}{4\pi}}\,F_{l0}(Q^2),
\end{equation}
so that we get at $Q^2=0$
\begin{displaymath}
G_{El}(0)=\frac{(2M)^l}{e}\,\frac{1}{(l!)^2}\int\ud^3r\,C_l(\vec{r})\,j^0(\vec{r}).
\end{displaymath}
The $l^\textrm{th}$ electric moment $Q_l$ \cite{moments} in (natural) unit of $e/M^l$ is therefore given by
\begin{equation}\label{normE}
Q_l\equiv \int\ud^3r\,C_l(\vec{r})\,j^0(\vec{r})=\frac{(l!)^2}{2^l}\,G_{El}(0).
\end{equation}

The multipole expansion \eqref{multdec2} of the charge density $\rho(\vec{r})=j^0(\vec{r})$ in terms of the multipole form factors \eqref{Edef} takes the form
\begin{equation}\label{Edec}
j^0(\vec{q})=e\sum_{l=0}^{+\infty}\,i^l\,\tau^{l/2}\,\frac{1}{\widetilde{C}_{2l-1}^{l-1}}\,G_{El}(Q^2)\,\sqrt{\frac{4\pi}{2l+1}}\,Y_{l0}(\Omega_q),
\end{equation}
where we have defined\footnote{We remind that by definition $(-1)!!=1$.}
\begin{equation}
\widetilde{C}_n^k\equiv\left\{\begin{array}{cl} \frac{n!!}{k!!\,(n-k)!!},&n\geq k\geq -1\\ 0,& \textrm{otherwise}\end{array}\right..
\end{equation}

\subsection{Magnetic Multipoles}

Magnetic multipoles are obtained from the magnetic density $\rho(\vec{r})=\vec{\nabla}\cdot\left(\vec{j}(\vec{r})\times\vec{r}\right)$. We define the magnetic multipole form factors $G_{Ml}(Q^2)$ as follows
\begin{equation}\label{Mdef}
G_{Ml}(Q^2)\equiv(-i)^l\,\frac{(2l-1)!!}{(l+1)!}\,\frac{(2M)^l}{e}\,\sqrt{\frac{2l+1}{4\pi}}\,F_{l0}(Q^2),
\end{equation}
so that we get at $Q^2=0$
\begin{displaymath}
G_{Ml}(0)=\frac{(2M)^l}{e}\frac{1}{(l!)^2}\int\ud^3r\,C_l(\vec{r})\,\frac{\vec{\nabla}\cdot\left(\vec{j}(\vec{r})\times\vec{r}\right)}{l+1}.
\end{displaymath}
The $l^\textrm{th}$ magnetic moment $\mu_l$ \cite{moments} in (natural) unit of $e/2M^l$ is therefore given by
\begin{equation}\label{normM}
\mu_l\equiv \int\ud^3r\,C_l(\vec{r})\,\frac{\vec{\nabla}\cdot\left(\vec{j}(\vec{r})\times\vec{r}\right)}{l+1}=\frac{(l!)^2}{2^{l-1}}\,G_{Ml}(0).
\end{equation}

The multipole expansion \eqref{multdec2} of the magnetic density $\rho(\vec{r})=\vec{\nabla}\cdot\left(\vec{j}(\vec{r})\times\vec{r}\right)$ in terms of the multipole form factors \eqref{Mdef} takes the form
\begin{equation}\label{Mdec}
\vec{\nabla}\cdot\left(\vec{j}(\vec{q})\times\vec{q}\right)=e\sum_{l=0}^{+\infty}\,i^l\,\tau^{l/2}\,\frac{(l+1)}{\widetilde{C}_{2l-1}^{l-1}}\,G_{Ml}(Q^2)\,Y_{l0}(\Omega_q).
\end{equation}

\subsection{Multipoles and Particles}

A particle of spin $j$ respecting parity symmetry has only \emph{even} electric multipoles and \emph{odd} magnetic multipoles \cite{ind}. Moreover, the total number of multipole form factors is equal to the total number of covariant vertex functions, namely $2j+1$. The decompositions \eqref{Edec} and \eqref{Mdec} take therefore the form
\begin{align}
j^0(\vec{q})&=\,e\mathop{\sum_{l=0}^{2j}}_{l\textrm{
even}}\,i^l\,\tau^{l/2}\,\frac{1}{\widetilde{C}_{2l-1}^{l-1}}\,G_{El}(Q^2)\,Y_{l0}(\Omega_q),\label{E}\\
\vec{\nabla}\cdot\left(\vec{j}(\vec{q})\times\vec{q}\right)&=\,e\mathop{\sum_{l=0}^{2j}}_{l\textrm{
odd}}\,i^l\,\tau^{l/2}\,\frac{(l+1)}{\widetilde{C}_{2l-1}^{l-1}}\,G_{Ml}(Q^2)\,Y_{l0}(\Omega_q),\label{M}
\end{align}
or more explicitly
\begin{displaymath}
\begin{split}
j^0(\vec{q})&=\,e\left[G_{E0}(Q^2)\,\sqrt{4\pi}\,Y_{00}(\Omega_q)-\frac{2}{3}\,\tau\,
G_{E2}(Q^2)\,\sqrt{\frac{4\pi}{5}}\,Y_{20}(\Omega_q)+\cdots\right],\\
\vec{\nabla}\cdot\left(\vec{j}(\vec{q})\times\vec{q}\right)&=\,e\,2i\sqrt{\tau}\left[G_{M1}(Q^2)\,\sqrt{\frac{4\pi}{3}}\,Y_{10}(\Omega_q)-\frac{4}{5}\,\tau
\,G_{M3}(Q^2)\,\sqrt{\frac{4\pi}{7}}\,Y_{30}(\Omega_q)+\cdots\right].
\end{split}
\end{displaymath}
Note that these decompositions do not depend on the particle spin. The latter just tells us what is the maximum allowed multipole in these series. 

In the next section, we will establish a direct relation between covariant vertex functions $F_k(Q^2)$ and electromagnetic multipole form factors $G_{El}(Q^2)$ and $G_{Ml}(Q^2)$. In order to make our job easier in establishing this connection, we will expand the spherical harmonics\footnote{For the sake of clarity, we omit the index $q$ attached to the angles, since we won't refer anymore to configuration space.} $Y_{l0}(\Omega)$ in terms of $\sin\theta$ and $\cos\theta$ \cite{angular}. For even values of $l$, the expansion reads
\begin{equation}\label{evenl}
Y_{l0}(\Omega)=\sqrt{\frac{2l+1}{4\pi}}\,\mathop{\sum_{s=0}^l}_{s\textrm{
even}}(-1)^{s/2}\,\widetilde{C}_l^s\,\widetilde{C}_{l+s-1}^{l-1}\,\sin^s\theta,
\end{equation}
while for odd values of $l$, it reads
\begin{equation}\label{oddl}
Y_{l0}(\Omega)=\sqrt{\frac{2l+1}{4\pi}}\,\cos\theta\mathop{\sum_{s=0}^{l-1}}_{s\textrm{
even}}(-1)^{s/2}\,\widetilde{C}_{l-1}^{l-s-1}\,\widetilde{C}_{l+s}^s\,\sin^s\theta.
\end{equation}

The expansion of the charge \eqref{E} and magnetic \eqref{M} densities becomes after insertion of \eqref{evenl} and \eqref{oddl}, respectively, and a few algebraic manipulations
\begin{align}
j^0(\vec{q})&=\,e\sum_{t=0}^n\sum_{m=t}^n(-1)^{m+t}\,\tau^m\,\frac{\left(C_m^t\right)^2}{\widetilde{C}_{4m-1}^{2m+2t-1}}\left(\sin^2\theta\right)^tG_{E2m}(Q^2),\label{Eprim}\\
\vec{\nabla}\cdot\left(\vec{j}(\vec{q})\times\vec{q}\right)&=\,e\,2i\sqrt{\tau}\,\cos\theta\sum_{t=0}^n\sum_{m=t}^n(-1)^{m+t}\left(1-\delta_{j,m}\right)\tau^m\,(m+1)\,\frac{\left(C_m^t\right)^2}{\widetilde{C}_{4m+1}^{2m+2t+1}}\left(\sin^2\theta\right)^tG_{M2m+1}(Q^2),\label{Mprim}
\end{align}
where $j=n$ for integer spin and $j=n+\frac{1}{2}$ for half-integer spin.

As a closing remark for this section, we would like to emphasize that \emph{only} $G_{E0}(0)$, $G_{M1}(0)$ and $G_{E2}(0)$ can directly be interpreted as electromagnetic moments. For $l>2$, the proportionality factor between electromagnetic moments and multipole form factors at $Q^2=0$ differs from unity, as one can see from eqs. \eqref{normE} and \eqref{normM}. 

\section{Breit Frame}\label{BF}

We want now to relate the covariant vertex functions $F_k(Q^2)$ of Section \ref{current} to the multipole form factors $G_{El}(Q^2)$ and $G_{Ml}(Q^2)$ of Section \ref{decomposition}. We follow the common usage by considering the so-called \emph{Breit frame} to connect the classical electromagnetic current $j^\mu$ of Section \ref{decomposition} with the quantum-mechanical one $J^\mu$ of Section \ref{current}
\begin{equation}\label{lien}
j^\mu(\vec{q})\equiv\frac{e}{2M}\,\langle
p',j|J^\mu_\textrm{EM}(0)|p,j\rangle=\frac{e}{2M}\,J^\mu_B.
\end{equation} 

The Breit frame (or brick-wall frame) is the frame where no energy
is transferred to the system by the photon\footnote{The reader might be worried by the fact that the definition of momentum $Q$ is different in Section \ref{current} compared to Section \ref{decomposition}. Covariant vertex functions and multipole form factors are however related in the Breit frame, where both definitions do actually match $Q^2\equiv -q^2=\vec{q}^{\,2}$.}
\begin{equation}
\begin{split}
q^\mu&=\,(0,\vec{q}),\\
p^\mu&=\,(p_0,-\frac{\vec{q}}{2}),\\
p'^\mu&=\,(p_0,\frac{\vec{q}}{2}),\\
p_0&=\,\sqrt{M^2+\frac{|\vec{q}|^2}{4}}=M\sqrt{1+\tau}.
\end{split}
\end{equation}
In this particular frame, the current $J^\mu_B$ has a non-relativistic appearance once explicitly expressed in terms of the rest-frame polarization
vectors $\vec{\varepsilon}_\lambda$ and rest-frame spinors $\chi_\lambda$ \cite{vect}. Since we have identified the spin quantization axis with the $z$-axis, the rest-frame polarization vectors and spinors are given by
\begin{gather}
\vec{\varepsilon}_{\pm}=\frac{1}{\sqrt{2}}\left(\mp
1,-i,0\right),\qquad
\vec{\varepsilon}_0=\left(0,0,1\right),\\
\chi_+=\left(\begin{array}{c}1\\0\end{array}\right),\qquad
\chi_-=\left(\begin{array}{c}0\\1\end{array}\right).
\end{gather}
In the following, we will use standard polarization four-vectors
\begin{equation}
\varepsilon^\mu(p,\lambda)=\left(\frac{\vec{\varepsilon}_\lambda\cdot\vec{p}}{M},\vec{\varepsilon}_\lambda+\frac{\vec{p}\left(\vec{\varepsilon}_\lambda\cdot\vec{p}\right)}{M(p_0+M)}\right),
\end{equation}
and standard spinors in Dirac representation
\begin{equation}
u(p,\lambda)=\sqrt{p_0+M}\left(\begin{array}{c}1\\ \frac{\vec{p}\cdot\vec{\sigma}}{p_0+M}\end{array}\right)\chi_\lambda.
\end{equation}

\subsection{Bosonic Case}

Considering $\lambda=\lambda'=j$, the polarization tensor \eqref{poltensor} of the particle takes a very simple form 
\begin{equation}
\varepsilon_{\alpha_1\cdots\alpha_j}(p,j)=\prod_{l=1}^j\,\varepsilon_{\alpha_l}(p,+).
\end{equation}
Let us discuss separately the charge and magnetic densities.

\subsubsection{Electric Part}

Concerning the charge density of integer spin particles, we find that all the complexity reduces to three structures only 
\begin{equation}\label{Electric}
\begin{split}
\varepsilon^*(p',+)\cdot\varepsilon(p,+)&=\,-1-\tau\,\sin^2\theta,\\
\frac{\left[\varepsilon^*(p',+)\cdot q\right]\left[\varepsilon(p,+)\cdot q\right]}{2M^2}&=\left(1+\tau\right)\tau\,\sin^2\theta,\\
\varepsilon^0(p,+)\left[\varepsilon^*(p',+)\cdot q\right]-\varepsilon^{0*}(p',+)\left[\varepsilon(p,+)\cdot q\right]&=\,2p_0\,\tau\,\sin^2\theta,
\end{split}
\end{equation}
leading to the expression 
\begin{displaymath}
J^0_B=2M\,\sum_{t=0}^j\sum_{k=0}^t\left(1+\tau\right)^{k+\frac{1}{2}}C_{j-k}^{j-t}\left(\tau\,\sin^2\theta\right)^t\left[F_{2k+1}(Q^2)-\frac{1-\delta_{k,0}}{1+\tau}\,F_{2k}(Q^2)\right].
\end{displaymath}
By comparison of this expression with \eqref{Eprim} and using \eqref{lien}, we obtain the connection we were looking for
\begin{equation}\label{Eintcon}
\sum^j_{m=t}(-1)^{m+t}\,\tau^{m-t}\,\frac{\left(C_m^t\right)^2}{\widetilde{C}_{4m-1}^{2m+2t-1}}\,G_{E2m}(Q^2)=\sum_{k=0}^t\left(1+\tau\right)^{k+\frac{1}{2}}C_{j-k}^{j-t}\left[F_{2k+1}(Q^2)-\frac{1-\delta_{k,0}}{1+\tau}\,F_{2k}(Q^2)\right].
\end{equation}

To illustrate this, let us consider for example the case $j=1$. The explicit expression of \eqref{Eintcon} becomes
\begin{displaymath}
\begin{split}
G_{E0}(Q^2)-\frac{2}{3}\,\tau\,G_{E2}(Q^2)&=\,\sqrt{1+\tau}\,F_1(Q^2),\\
G_{E2}(Q^2)&=\,\sqrt{1+\tau}\left[F_1(Q^2)-F_2(Q^2)+(1+\tau)\,F_3(Q^2)\right],
\end{split}
\end{displaymath}
which coincides with the well-known expression for the electromagnetic interaction of vector particles \cite{vect}, provided that
\begin{displaymath}
G_{E0}(Q^2)=\sqrt{1+\tau}\,G_C(q^2),\qquad G_{E2}(Q^2)=\sqrt{1+\tau}\,G_Q(q^2).
\end{displaymath}
We remind that $\sqrt{1+\tau}=p_0/M$ in the Breit frame.

\subsubsection{Magnetic Part}

Concerning the magnetic density of integer spin particles, the first simplification is
\begin{equation}
\vec{P}\times\vec{q}=0,
\end{equation}
so that only covariant vertex functions with even value of $k$ will contribute. Moreover, we have in the Breit frame
\begin{equation}\label{Magnetic}
\begin{split}
\vec{\nabla}\cdot\left[\vec{\varepsilon}(p,\lambda)\times\vec{q}\right]&=\,0,\\
\left[\vec{\varepsilon}(p,\lambda)\times\vec{q}\right]\cdot\vec{\nabla}\left[\varepsilon(p,\lambda)\cdot q\right]&=\,0,\\
\left[\vec{\varepsilon}(p,\lambda)\times\vec{q}\right]\cdot\vec{\nabla}\left[\varepsilon^*(p',\lambda')\cdot q\right]&=\,-\frac{p_0}{M}\left(\vec{\varepsilon}^*_{\lambda'}\times\vec{\varepsilon}_\lambda\right)\cdot\vec{q},
\end{split}
\end{equation}
from which we obtain
\begin{equation}
\begin{split}
&\vec{\nabla}\cdot\left\{\left\{\vec{\varepsilon}(p,+)\left[\varepsilon^*(p',+)\cdot q\right]-\vec{\varepsilon}^*(p',+)\left[\varepsilon(p,+)\cdot q\right]\right\}\times\vec{q}\left(\frac{\left[\varepsilon^*(p',+)\cdot q\right]\left[\varepsilon(p,+)\cdot q\right]}{2M^2}\right)^kf(Q^2)\right\}\\
&\qquad=-2p_0\,(k+1)\,2i\sqrt{\tau}\,\cos\theta\left(\tau\,\sin^2\theta\right)^kf(Q^2).
\end{split}
\end{equation}
This leads to the desired expression
\begin{displaymath}
\vec{\nabla}\cdot\left(\vec{J}_B\times\vec{q}\right)=4iM\sqrt{\tau}\,\cos\theta\sum_{t=0}^{j-1}\sum_{k=0}^t\left(1+\tau\right)^{k+\frac{1}{2}} C_{j-k-1}^{j-t-1}\left(\tau\,\sin^2\theta\right)^t\left(t+1\right)\,F_{2k+2}(Q^2).
\end{displaymath}
By comparison of this expression with \eqref{Mprim} and using \eqref{lien}, we obtain the connection we were looking for
\begin{equation}\label{Mintcon}
\sum^{j-1}_{m=t}(-1)^{m+t}\,\tau^{m-t}\,\left(m+1\right)\,\frac{\left(C_m^t\right)^2}{\widetilde{C}_{4m+1}^{2m+2t+1}}\,G_{M2m+1}(Q^2)=\left(t+1\right)\sum_{k=0}^t\left(1+\tau\right)^{k+\frac{1}{2}}C_{j-k-1}^{j-t-1}\,F_{2k+2}(Q^2).
\end{equation}

To illustrate this, let us consider for example the case $j=1$. The explicit expression of \eqref{Mintcon} becomes
\begin{displaymath}
G_{M1}(Q^2)=\sqrt{1+\tau}\,F_2(Q^2),
\end{displaymath}
which coincides with the well-known expression for the electromagnetic interaction of vector particles \cite{vect}, provided that
\begin{displaymath}
G_{M1}(Q^2)=\sqrt{1+\tau}\,G_M(q^2).
\end{displaymath}
The factor $\sqrt{1+\tau}$ is present for any integer spin. One is free to define new multipole form factors without this inelegant factor, at the cost of having different multipole decompositions for fermions and bosons. Nevertheless, at $Q^2=0$ there is no difference between both definitions.

\subsubsection{Real Photon Point $Q^2=0$}

Our final aim is to determine the natural electromagnetic moments. Let us therefore concentrate on the real photon point $Q^2=0$. In this limit, the relations \eqref{Eintcon} and \eqref{Mintcon} reduce to
\begin{subequations}\label{Dirint}
\begin{align}
G_{E2m}(0)&=\sum_{k=0}^mC_{j-k}^{j-m}\left[F_{2k+1}(0)-(1-\delta_{k,0})\,F_{2k}(0)\right],\\
G_{M2m+1}(0)&=\sum_{k=0}^m C_{j-k-1}^{j-m-1}\,F_{2k+2}(0).
\end{align}
\end{subequations}

These relations can be inverted to express $F_k(0)$ in terms of $G_{El}(0)$ and $G_{Ml}(0)$
\begin{subequations}\label{Invint}
\begin{align}
F_{2k+1}(0)=&\,\sum_{l=0}^kC_{j-l}^{j-k}(-1)^{k-l}\left[G_{E2l}(0)+(1-\delta_{l,0})\,G_{M2l-1}(0)\right],\\
F_{2k+2}(0)=&\,\sum_{l=0}^k C_{j-l-1}^{j-k-l}(-1)^{k-l}\,G_{M2l+1}(0),
\end{align}
\end{subequations}
thanks to the following identity
\begin{equation}\label{identity}
\sum_{l=m}^kC_{j-l}^{j-k}\,C_{j-m}^{j-l}\,(-1)^{k-l}=\delta_{k,m},\qquad\qquad\forall j\geq k.
\end{equation}

\subsection{Fermionic Case}

We proceed with the fermionic case in a way completely anologous to the bosonic one. Considering $\lambda=\lambda'=j=n+\frac{1}{2}$, the polarization spin-tensor \eqref{spintensor} of the particle takes a very simple form 
\begin{equation}
u_{\alpha_1\cdots\alpha_n}(p,j)=u(p,+)\prod_{l=1}^n\,\varepsilon_{\alpha_l}(p,+).
\end{equation}
Let us discuss separately the charge and magnetic densities.

\subsubsection{Electric Part}

Concerning the charge density of half-integer spin particles, we find that all the complexity reduces to the three structures of the bosonic case \eqref{Electric} \emph{plus} two new ones
\begin{equation}
\begin{split}
\bar u(p',+)\,\gamma^0\,u(p,+)&=\,2M,\\
\bar u(p',+)\,\frac{i\sigma^{0\nu}q_\nu}{2M}\,u(p,+)&=\,-2M\tau,
\end{split}
\end{equation}
leading to the expression 
\begin{displaymath}
J^0_B=2M\sum_{t=0}^n\sum_{k=0}^t\left(1+\tau\right)^k C_{n-k}^{n-t}\left(\tau\,\sin^2\theta\right)^t\left[F_{2k+1}(Q^2)-\tau\,F_{2k+2}(Q^2)\right].
\end{displaymath}
By comparison of this expression with \eqref{Eprim} and using \eqref{lien}, we obtain the connection we were looking for
\begin{equation}\label{Ehalfintcon}
\sum_{m=t}^n(-1)^{m+t}\,\tau^{m-t}\,\frac{(C_m^t)^2}{\widetilde{C}_{4m-1}^{2m+2t-1}}\,G_{E2t}(Q^2)=\sum_{k=0}^t\left(1+\tau\right)^k C_{n-k}^{n-t}\left[F_{2k+1}(Q^2)-\tau\,F_{2k+2}(Q^2)\right].
\end{equation}

To illustrate this, we will consider two examples. Let us consider first the case $j=1/2$. The explicit expression of \eqref{Ehalfintcon} becomes
\begin{displaymath}
G_{E0}(Q^2)=F_1(Q^2)-\tau\,F_2(Q^2),
\end{displaymath}
which coincides with the well-known expression for the electromagnetic interaction of spin-$1/2$ particles. Our electric monopole form factor $G_{E0}(Q^2)$ is nothing more than the Sachs electric form factor $G_E(Q^2)$. Let us consider now the case $j=3/2$. The explicit expression of \eqref{Ehalfintcon} becomes
\begin{displaymath}
\begin{split}
G_{E0}(Q^2)-\frac{2}{3}\,\tau\,G_{E2}(Q^2)&=\,F_1(Q^2)-\tau\,F_2(Q^2),\\
G_{E2}(Q^2)&=\,F_1(Q^2)-\tau\,F_2(Q^2)+\left(1+\tau\right)\left[F_3(Q^2)-\tau\,F_4(Q^2)\right],
\end{split}
\end{displaymath}
which coincides with the well-known expression for the electromagnetic interaction of spin-$3/2$ particles \cite{vdh}. From these expressions, it is obvious that our definition of covariant vertex functions $F_k(Q^2)$ is more economical than in \cite{vdh}, by avoiding spurious $(-\frac{1}{2})$ factors.

\subsubsection{Magnetic Part}

Concerning the magnetic density of half-integer spin particles, we have in addition to \eqref{Electric} 
\begin{equation}
\bar u(p',+)\,\gamma^k\,u(p,+)=\bar u(p',+)\,\frac{i\sigma^{k\nu}q_\nu}{2M}\,u(p,+)=\chi^\dag_+\,i\left(\vec{\sigma}\times\vec{q}\right)^k\chi_+,
\end{equation}
from which we obtain
\begin{equation} 
\begin{split}
&\vec{\nabla}\cdot\left\{\chi^\dag_+\left[i\left(\vec{\sigma}\times\vec{q}\right)\times\vec{q}\right]\chi_+\left[\frac{\left(\varepsilon'^*(+1)\cdot q\right)\left(\varepsilon(+1)\cdot q\right)}{2M^2}\right]^kf(Q^2)\right\}\\
&\qquad=4M\,(k+1)\,i\sqrt{\tau}\,\cos\theta\left(\tau\,\sin^2\theta\right)^kf(Q^2).
\end{split}
\end{equation}
This leads to the desired expression
\begin{displaymath}
\vec{\nabla}\cdot\left(\vec{J}_B\times\vec{q}\right)=4M\,i\sqrt{\tau}\,\cos\theta\sum_{t=0}^n\sum_{k=0}^t\left(1+\tau\right)^k C_{n-k}^{n-t}\left(\tau\,\sin^2\theta\right)^t\left(t+1\right)\left[F_{2k+1}(Q^2)+F_{2k+2}(Q^2)\right].
\end{displaymath}
By comparison of this expression with \eqref{Mprim} and using \eqref{lien}, we obtain the connection we were looking for
\begin{equation}\label{Mhalfintcon}
\sum_{m=t}^n(-1)^{m+t}\,\tau^{m-t}\,\left(m+1\right)\,\frac{\left(C_m^t\right)^2}{\widetilde{C}_{4m+1}^{2m+2t+1}}\,G_{M2t+1}(Q^2)=\left(t+1\right)\sum_{k=0}^t\left(1+\tau\right)^k C_{n-k}^{n-t}\left[F_{2k+1}(Q^2)+F_{2k+2}(Q^2)\right].
\end{equation}

To illustrate this, we will consider two examples. Let us consider first the case $j=1/2$. The explicit expression of \eqref{Mhalfintcon} becomes
\begin{displaymath}
G_{M1}(Q^2)=F_1(Q^2)+F_2(Q^2),
\end{displaymath}
which coincides with the well-known expression for the electromagnetic interaction of spin-$1/2$ particles. Our magnetic dipole form factor $G_{M1}(Q^2)$ is nothing more than the Sachs magnetic form factor $G_M(Q^2)$. Let us consider now the case $j=3/2$. The explicit expression of \eqref{Mhalfintcon} becomes
\begin{displaymath}
\begin{split}
G_{M1}(Q^2)-\frac{4}{5}\,\tau\,G_{M3}(Q^2)&=\,F_1(Q^2)+F_2(Q^2),\\
G_{M3}(Q^2)&=\,F_1(Q^2)+F_2(Q^2)+\left(1+\tau\right)\left[F_3(Q^2)+F_4(Q^2)\right],
\end{split}
\end{displaymath}
which coincides with the well-known expression for the electromagnetic interaction of spin-$3/2$ particles \cite{vdh}. Once more, one can see that our definition of covariant vertex functions $F_k(Q^2)$ is more economical than in \cite{vdh}, by avoiding spurious $(-\frac{1}{2})$ factors. We would like also to remind that $G_{M3}(0)$ does \emph{not} correspond to the value of the magnetic octupole, contrarily to what can be found in the literature, but represents one ninth of its value.

\subsubsection{Real Photon Point $Q^2=0$}

Let us now concentrate on the real photon point $Q^2=0$. In this limit, the relations \eqref{Ehalfintcon} and \eqref{Mhalfintcon} reduce to
\begin{subequations}\label{Dirhalfint}
\begin{align}
G_{E2m}(0)&=\sum_{k=0}^mC_{n-k}^{n-m}\,F_{2k+1}(0),\\
G_{M2m+1}(0)&=\sum_{k=0}^m C_{n-k}^{n-m}\left[F_{2k+1}(0)+F_{2k+2}(0)\right].
\end{align}
\end{subequations}

These relations can be inverted to express $F_k(0)$ in terms of $G_{El}(0)$ and $G_{Ml}(0)$
\begin{subequations}\label{Invhalfint}
\begin{align}
F_{2k+1}(0)=&\,\sum_{l=0}^k C_{n-l}^{n-k}\,(-1)^{k-l}\,G_{E2l}(0),\\
F_{2k+2}(0)=&\,\sum_{l=0}^k C_{n-l}^{n-k}\,(-1)^{k-l}\left[G_{M2l+1}(0)-G_{E2l}(0)\right],
\end{align}
\end{subequations}
thanks to the identity \eqref{identity}.

\section{Conclusion}

In this paper, we have presented the first part of our results. We have proposed transparent expressions for the arbitrary-spin electromagnetic current in terms of covariant vertex functions. Performing an explicit multipole decomposition, we have defined generally the multipole form factors and worked out their relation with the electromagnetic moments. In the Breit frame, we were able to derive the general relation between the covariant vertex functions and multipole form factors. We naturally recover the low-spin cases studied so far.

Beside the fact that the steps explained in this paper are necessary for our aim of obtaining the natural electromagnetic moments, the results presented here will be relevant for other studies. For example, based on our decomposition in terms of covariant vertex functions, it should be in principle possible to determine within Lattice QCD the structure of high-spin resonances. It should also be stressed that our decomposition is very economical since it avoids irrelevant factors in the expressions. Moreover, it has been shown explicitly that multipole form factors at $Q^2=0$ are in general \emph{not} equal to the electromagnetic moments. The identification is valid only up to quadrupoles.

\subsection*{Acknowledgements}

The author is grateful to M. Vanderhaeghen, V. Pascalutsa and T. Ledwig for numerous enlightening discussions and comments.

\end{document}